\documentclass[]{iopart}
\usepackage{iopams}
\usepackage{graphicx}
\begin{document}

\title{Dynamical screening in a quark gluon plasma}

\author{Munshi G. Mustafa\dag\ \footnote[3]{mustafa@theory.saha.ernet.in} 
, Purnendu Chakraborty\dag\ , 
 and Markus H. Thoma\ddag}

\address{\dag\ Theory Division, Saha Institute of Nuclear Physics,
  1/AF Bidhannagar, Kolkata, India. }

\address{\ddag Centre for Interdisciplinary Plasma Science,
Max-Planck-Institut f\"ur extraterrestrische Physik,
P.O. Box 1312, 85741 Garching, Germany}

\begin{abstract}
We calculate the screening potential of a fast parton moving through 
a quark-gluon plasma in the framework of semi-classical transport theory. 
We found an anisotropic potential showing a minimum in the direction of 
the parton velocity. Possible consequences of this potential on binary 
states in a quark-gluon plasma are discussed.

\end{abstract}

%\pacs{12.38.Mh}

%\submitto{\JPG}

%\maketitle

%\vspace{.5cm}
 
Screening of charges in a plasma is an important collective 
effect in plasma physics. In a classical isotropic 
and homogeneous plasma the screening potential of a point-like
test charge $Q$ at rest  is modified from Coulomb potential into a
%Debye-H\"uckel or
Yukawa potential \cite{LL} as
\begin{equation}
\phi (r) = \frac{Q}{r} \exp (-m_D r)
\label{e1}
\end{equation}
with the Debye mass (inverse screening length) $m_D$ ($\hbar = c = k_B =1$). 
In the quark-gluon plasma the Debye mass of a chromoelectric charge follows
from the static limit of the longitudinal polarization tensor which in the 
high temperature limit and one loop order is given by \cite{Kajantie,Silin,Klimov, 
Weldon},
\begin{equation}
\Pi_{00} (\omega =0, k) = -m_D^2 = -g^2T^2 \left (1+\frac{n_f}{6}\right ),
\label{e2} 
\end{equation}
where $g$ is the strong coupling constant and  $n_f$ the number of light 
quark flavors in the QGP with $m_q \ll T$.

The modification of the confinement potential below the critical temperature
into a Yukawa potential above the critical temperature
might have important consequences for the discovery 
of the QGP in relativistic heavy-ion collisions. Bound states of
heavy quarks, in particular the $J/\psi$ meson, which are produced in the
initial hard scattering processes of the collision, will be dissociated in 
the QGP due to screening of the quark potential and break-up by energetic
gluons \cite{Patra}. Hence the suppression
of $J/\psi$ mesons have been proposed as one of the most promising
signatures for the QGP formation \cite{Matsui}. On the other hand, the
formation of colored bound states, {\textit e.g.}, $qq$, $\bar q\bar q$, $gg$, 
of partons 
at rest has also been claimed~\cite{Shuryak} above the critical
temperature (2$T_c$ - 3$T_c$) by analyzing lattice data,
indicating that the plasma behaves as a strongly coupled quark-gluon plasma.

%Indeed, a suppression
%of $J/\psi$ mesons has been observed experimentally \cite{exper} and
%interpreted as a strong indication for the QGP formation in relativistic
%heavy-ion collisions \cite{Blaizot}.

Earlier in most  calculations of the screening potential in the QGP, the
test charge was assumed to be at rest. However, quarks and gluons
coming from initial hard processes receive a transverse momentum which causes
them to propagate through the QGP \cite{jet}. In addition, hydrodynamical
models predict a radial outward flow in the fireball \cite{flow}. Hence,
it is of great interest to estimate the screening potential of a parton
moving~\cite{munshi} relatively to the QGP. Chu and Matsui \cite{Chu} have used
the Vlasov equation to investigate dynamic Debye screening for a heavy
quark-antiquark pair traversing a quark-gluon plasma. They found that the 
screening potential becomes strongly anisotropic.

%In the case of a non-relativistic plasma the 
The screening potential of a moving charge $Q$ with velocity $v$ follows from
the linearized Vlasov and Poisson equations as \cite{Ichimaru,Spatschek}
\begin{equation}
\phi({\vec r}, t; {\vec v}) = \frac{Q}{2\pi^2} \int d^3k
\frac{\exp{[-i{\vec k}\cdot ({\vec r}-{\vec v}t)]}}
{k^2 {\rm Re}[\epsilon_l (\omega={\vec k}\cdot {\vec v}, k)]}.
\label{e4}
\end{equation}
%to the classical approximation. For instance, it is related
The dielectric function following from the semi-classical
Vlasov equation describing a collisionless plasma is related to
the high-temperature limit of the polarization tensor.
For example, the longitudinal dielectric function following from 
the Vlasov equation is given by \cite{Silin,Klimov,Weldon}
\begin{eqnarray}
\epsilon_l (\omega ,k) = 1-\frac{\Pi_{00}(\omega ,k)}{k^2}
= 1+\frac{m_D^2}{k^2} \left (1-\frac{\omega}{2k} \ln \frac {\omega +k}
{\omega -k}\right ),
\label{e3}
\end{eqnarray}
where the only non-classical inputs are 
Fermi and Bose distributions instead of the Boltzmann distribution. 
The gluon self-energies derived within the hard thermal loop 
approximation~\cite{Braaten} have been shown to be gauge invariant 
and the dielectric functions obtained from these are therefore 
also gauge invariant.

%It is easy to show that this expression reduces to the Yukawa potential in 
%the case of small velocities, $v=|{\vec v}\, |\ll v_{th}$, 
%where $v_{th}$ is the thermal 
%velocity of the plasma particles. In the opposite case, $v\gg v_{th}$,
%the Coulomb potential is recovered since a screening charge cloud cannot be
%formed for fast particles.
%The above equation (\ref{e4}) also holds in the case of a relativistic plasma.
%We only have to use the relativistic expression (\ref{e3}) for the 
%longitudinal dielectric function. 
For small velocities, $v \rightarrow 0$, 
i.e. $\omega \ll k$, we obtain
\begin{equation}
\epsilon_l (\omega \ll k) = 1+\frac{m_D^2}{k^2},
\label{e5}
\end{equation}
from which again the (shifted) Yukawa potential results \cite{Spatschek}
\begin{equation}
\phi({\vec r}, t; {\vec v}) = \frac{Q}{|{\vec r}-{\vec v}t|} \,
\exp (-m_D |{\vec r}-{\vec v}t|).
\label{e6}
\end{equation}
%It should be noted that the opposite limit $v\gg v_{th}$, 
%leading to a Coulomb potential in the non-relativistic case,
%cannot be realized in an ultrarelativistic plasma because the thermal 
%velocity of the plasma particles is given by the speed of light, $v_{th}=c=1$.

In the general case, for parton velocities $v$ 
between 0 and 1, we have to 
solve (\ref{e4}) together with (\ref{e3}) numerically. Since the potential
is not isotropic anymore due to the velocity vector ${\vec v}$,
we will restrict ourselves only to two cases, ${\vec r}$ parallel to ${\vec v}$
and ${\vec r}$ perpendicular to ${\vec v}$, i.e., for illustration
we consider~\cite{munshi} the screening potential only in the direction 
of the moving parton or perpendicular to it. 

In Fig.1 the screening potential
$\phi /Q$ in ${\vec v}$-direction is shown as a function of $r'=r-vt$, 
where $r=|{\vec r}|$, between 0 and 6 fm for various velocities. 
For illustration we have chosen a strong
fine structure constant $\alpha_s =g^2/(4\pi)=0.3$, 
a temperature $T=0.25$ GeV, and the number of quark flavors $n_f=2$. 
The shifted potentials depend only on $v$ and
not on $t$ as it should be the case in a homogeneous and isotropic plasma. 
For $r'<1$ fm one observes that the fall-off of the potential
is stronger than for a parton at rest.
The reason for this behavior is the fact that there is a stronger 
screening in the direction 
of the moving parton due to an enhancement of the particle density 
in the rest frame of the moving parton.

In addition, a minimum in the screening potential at $r'>1$ fm shows up. For 
example, for $v=0.8$ this minimum is at about 1.5 fm with a depth 
of about 8 MeV. 
%The occurrence of a minimum in the potential in the 
%direction of the velocity is also observed in so-called complex plasmas. 
%Complex plasmas are classical, low-temperature plasmas 
%containing particles with a diameter of a few microns \cite{Merlino}. These 
%particles are charged in the plasma by collecting electrons on their surface.
%In the presence of an ion flow the positively charged ions are deflected 
%by the
%microparticles, leading to an anisotropic distortion of the Debye sphere by
%an enhancement of the ion density in front of the microparticle. This 
%positive charge cloud leads to an attraction between microparticles
%in the direction of the ion flow and the formation of string like 
%structures \cite{Shukla}, observed in experiments.
 A minimum in the screening potential is also known from non-relativistic, 
complex
plasmas, where an attractive potential even between equal charges
can be found if the finite extension of the charges is considered 
\cite{Tsytovich}. 
A similar screening potential was found for a color charge at rest
in Ref.\cite{Gale}, where a polarization tensor beyond the high-temperature 
limit was used. 
However, this approach has its limitation as a gauge dependent and incomplete 
(within the order
of the coupling constant) approximation for the polarization tensor was used \cite{Braaten}. 
Obviously, a minimum in the interparticle potential in a relativistic
or non-relativistic plasma is a general feature if one goes beyond the Debye-H\"uckel
approximation by either taking quantum effects, finite velocities, or finite sizes of the particles
into account.
 
We also note that Chu and Matsui \cite{Chu} did not report the existence 
of a minimum in the potential
of a quark traversing the QGP. However, in their Fig.1(d) a negative value of the potential of
a fast quark ($v=0.9$) in the direction of the quark velocity is shown. Since the potential has 
to tend to zero for large distances, this implies a minimum in the screening potential.
This minimum was not found because the potential was plotted only for the limited range
$0<r<1/m_d\simeq 0.35$ fm for our choice of the parameters.

The minimum could give rise to bound states, e.g., of diquarks, if thermal fluctuations do not 
destroy them. The two-body potential, associated with the dipole fields 
created by two test charges $Q_1$ and $Q_2$ at ${\vec r_1}$ and ${\vec r_2}$
with velocities ${\vec v_1}$ and ${\vec v_2}$, can be written as 
\begin{eqnarray}
\Phi ({\vec r_1}-{\vec r_2}, {\vec v_1-\vec v_2}, t) &=& 
%\nonumber \\
\frac{Q_1Q_2}{4\pi^2}\> \int d^3k \>
\Biggl \{ \frac{e^{[i{\vec k}\cdot (-({\vec r_1}-{\vec r_2})-({\vec v_1}-{\vec v_2})t)]}}
{k^2 {\rm Re}[\epsilon_l (\omega={\vec k}\cdot {\vec v_1}, k)]}\nonumber \\
&&+\frac{e^{[i{\vec k}\cdot (({\vec r_1}-{\vec r_2})-({\vec v_1}-{\vec v_2})t)]}}
{k^2 {\rm Re}[\epsilon_l (\omega={\vec k}\cdot {\vec v_2}, k)]}\Biggr \}.
\label{e7}
\end{eqnarray}
For comoving quarks, ${\vec v_1} = {\vec v_2}$, this two-body potential reduces
to the one-body potential (\ref{e4}), showing the attraction between the quarks
which could give rise to a bound state.
Colored bound states, e.g., diquarks, of partons at rest have also been 
claimed by analyzing lattice data \cite{Shuryak}. For a 
comoving diquark system,
where the interaction between two color charges $Q_1Q_2>0$, the two body
potential shows an attractive minimum and may therefore
indicate a possible appearance of
a diquark bound state as compared to partons at rest where the two body 
potential is only repulsive.  On the other hand,
for a quark-antiquark system, where $Q_1Q_2<0$, the two-body potential 
is inverted, showing a maximum. This may lead to short living mesonic 
resonances and an enhancement of the attraction between quarks and 
antiquarks of mesonic states moving through the QGP.   

%%%%%%%%%%%%%%%%%%% figure 1 & 2 together %%%%%%%%%%%%%%%%%%%%%%%%%
%\vspace{-0.35in}
\begin{figure}[h]
%\epsfxsize=3.8in
%\hspace{-0.2in}\epsfbox{Fig1.ps}
%\vspace{-0.8in}

\begin{minipage}[bl]{6.5cm}
\includegraphics[width=6cm,height=6cm]{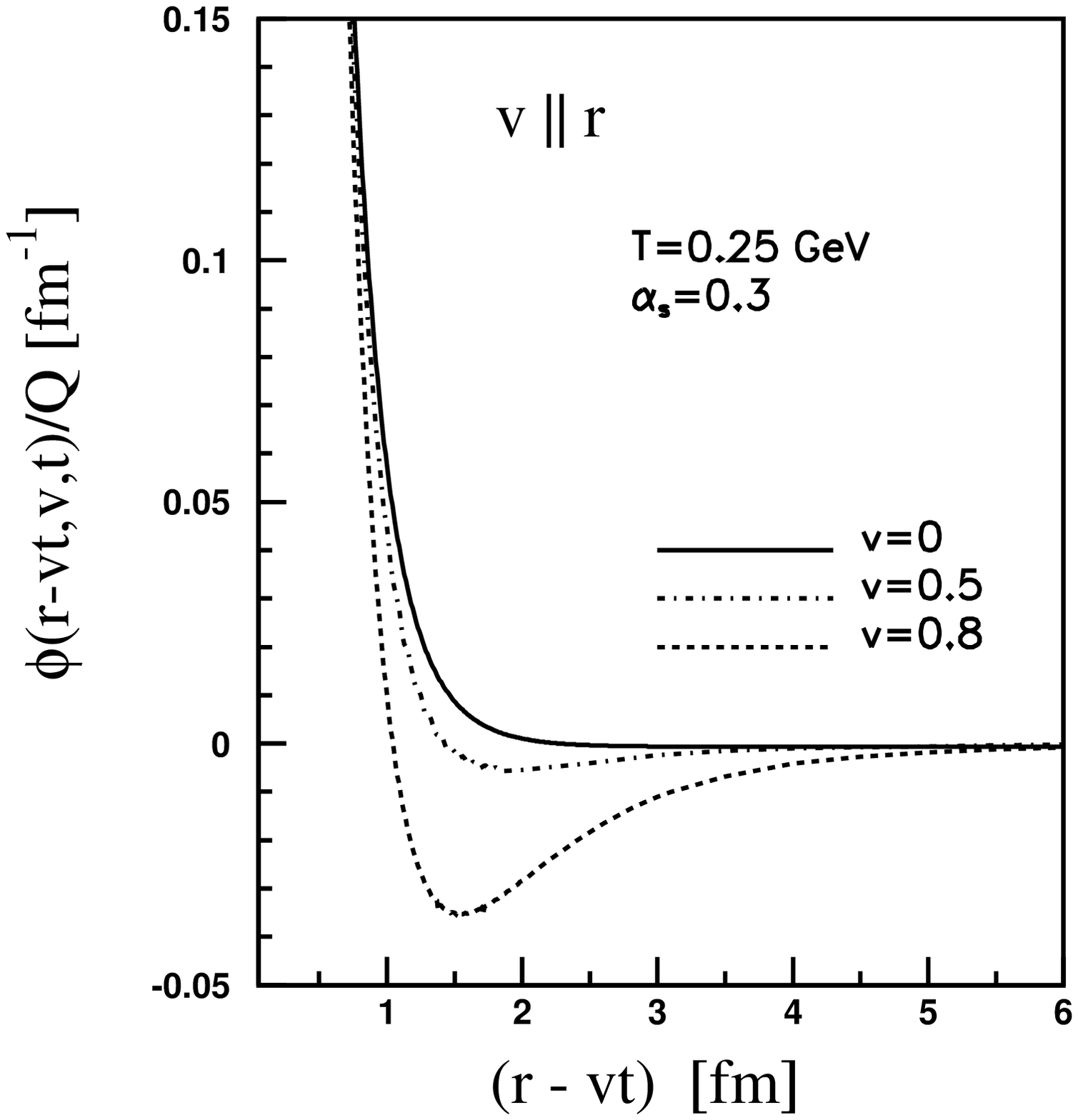}
\caption{\label{plot1}\noindent 
{Screening potential parallel to the velocity of the 
moving
parton in a QGP as a function of $r-vt$ (0 to 6 fm) for $v=0$, 0.5, and 0.8.
}}  
\end{minipage}
\hfill 
\begin{minipage}[bl]{6.5cm}
\includegraphics[width=6cm,height=6cm]{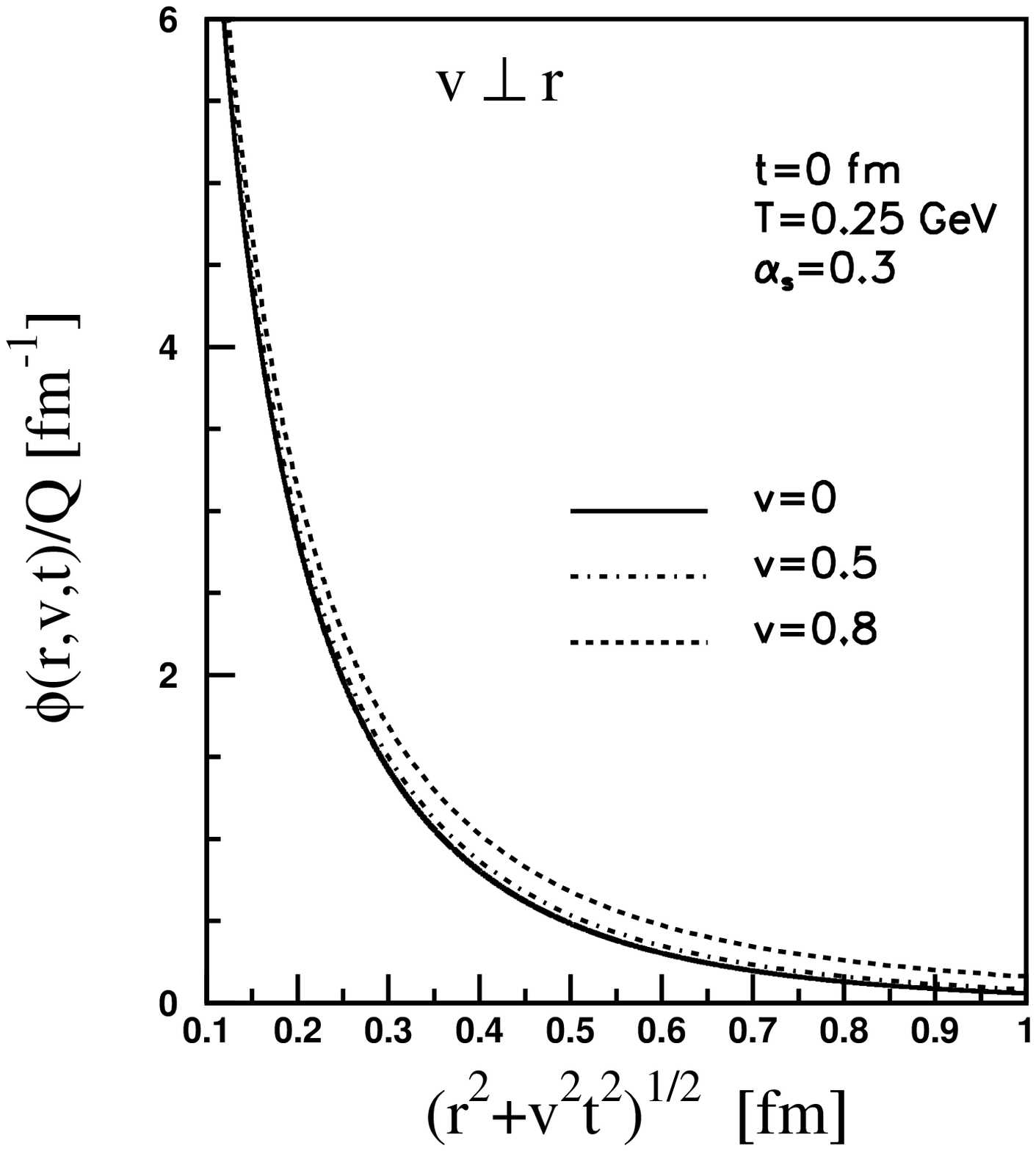}
\caption{\label{plot2}
Screening potential perpendicular to the velocity of the moving
parton in a QGP as a function of $| {\vec r} -{\vec v}t|$ (0.1 to 1 fm) for 
$v=0$, 0.5, and 0.8.}
\end{minipage}
%\noindent{Fig.1: Screening potential parallel to the velocity of the moving
%parton in a QGP as a function of $r-vt$ (0 to 6 fm) for $v=0$, 0.5, and 0.8.}
\end{figure}
%%%%%%%%%%%%%%%%%%% end of figure 1 %%%%%%%%%%%%%%%%%%%%%%%%%

%The details of the potential, e.g., the depth of the minimum,
%depend on the choice of the parameters, such as the coupling constant. For a 
%value $\alpha_s =0.3$, as it is typical for the temperature reachable in
%heavy-ion collisions, the semi-classical approach corresponding to the
%weak coupling limit might not be reliable. Quantum effects and collisions
%between plasma particles
%are important at those temperatures and will change the dielectric functions 
%and dispersion
%relations. Within the transport theoretical approach collisions can be 
%considered,
%for example by using the relaxation time approximation \cite{Carrington}. 

%Non-abelian effects (beyond color factors, e.g., in the Debye mass)
%will be important at realistic temperatures. Unfortunately they cannot be 
%treated by the methods used here and are therefore beyond the scope of this 
%work. However, as we 
%discussed above, also in a complex plasma which is a strongly coupled plasma 
%as it is also the case for the QGP close to
%the critical temperature, there is an attraction between the microparticles 
%in the presence of an ion flow. This appears to be a general feature of 
%weakly as well as strongly coupled plasmas.
%Therefore we do not expect a qualitative change of the screening potential 
%due to non-perturbative and non-abelian effects. 

The results for ${\vec r}$ perpendicular to ${\vec v}$ are shown in Fig.2,
where the potential is shown as a function of $|{\vec r} - {\vec v}t|=\sqrt{r^2+v^2t^2}$
between 0.1 and 1 fm.
Here we consider only the case $t=0$ since at $t>0$ and $v>0$ there is no singularity
in the potential due to $\sqrt{r^2+v^2t^2}>0$ for all $r$. Hence the potential
is cut-off artificially at small distances if plotted as a function of $\sqrt{r^2+v^2t^2}$.
In contrast to the parallel case (Fig.1)
the fall-off of the potential at larger values of $v$ is less steep, i.e. the screening is
reduced as it is expected since the formation of the screening cloud is suppressed at large 
velocities. Also no minimum in the potential is found.

%%%%%%%%%%%%%%%%%%% figure 2  here %%%%%%%%%%%%%%%%%%%%%%%%%
%\vspace{-0.35in}
%\begin{figure}
%\epsfxsize=3.8in
%\hspace{-0.2in}\epsfbox{Fig2.ps}
%\vspace{-0.8in}

%\noindent{Fig.2: Screening potential perpendicular to the velocity of 
%the moving
%parton in a QGP as a function of $| {\vec r} -{\vec v}t|$ (0.1 to 1 fm) for 
%$v=0$, 0.5, and 0.8.}
%\end{figure}
%%%%%%%%%%%%%%%%%%% end of figure 2 %%%%%%%%%%%%%%%%%%%%%%%%%

Summarizing, we have calculated the screening potential of a fast color charge
moving through the QGP from semi-classical transport theory corresponding to the high-temperature 
limit. As in Ref.\cite{Chu} the potential is found to be strongly anisotropic.
The screening is reduced in the perpendicular direction of the moving
parton but increased in the direction of the moving parton, which may 
lead to a modification of the $J/\psi$ suppression. In addition,
we found a new feature of the screening potential of a fast parton in a QGP:
a minimum in the potential shows up which could give rise to bound states 
of, for example, diquarks if not destroyed by thermal fluctuations. 
For a quark-antiquark pair this minimum turns into a maximum which 
could cause short living mesonic resonances. Combining the effect of 
reduced screening in perpendicular direction and the presence of a maximum 
in parallel direction we expect a stronger binding of 
moving $J/\psi$ mesons with respect to the QGP than of $J/\psi$ mesons at rest.
The consequences, e.g, 
for the $J/\psi$ yield should be investigated in more detail using 
hydrodynamical models or
event generators for the space-time evolution of the fireball. Finally, let us note that our results also 
apply to other ultrarelativistic plasmas such as an electron-positron plasma in Supernova explosions.
In this case one simply has to replace the Debye mass $m_D$ by $eT$, where $e$ is the electron charge.

\end{document}